\documentclass[aps,twocolumn,pre,superscriptaddress,noeprint]{revtex4-1}

\usepackage{dcolumn}
\usepackage{amsmath}
\usepackage[dvipsnames,usenames]{color}
\usepackage{amssymb,amsbsy}
\usepackage{xr}
\usepackage[utf8]{inputenc}
\usepackage{xr-hyper}
\usepackage{nameref}
\usepackage{graphicx}
\usepackage{pslatex}
\usepackage{soul}

\newlength{\figurewidth}
\begin{document}

\title{Turing patterns mediated by network topology in homogeneous active systems}
\author{Sayat Mimar}
\affiliation{Department of Physics \& Astronomy, University of Rochester, Rochester, NY 14607, USA}
\author{Mariamo Mussa Juane}
\affiliation{Group of Nonlinear Physics. University of Santiago de Compostela, Santiago de Compostela 15782, Spain}
\author{Juyong Park}
\affiliation{Graduate School of Culture Technology, Korea Advanced Institute of Science and Technology, Daejon, 305-701, Korea}
\author{Alberto P. Mu\~nuzuri}
\affiliation{Group of Nonlinear Physics. University of Santiago de Compostela, Santiago de Compostela 15782, Spain}
\author{Gourab Ghoshal}
\affiliation{Department of Physics \& Astronomy, University of Rochester, Rochester, NY 14607, USA}

\begin{abstract}
	Mechanisms of pattern formation---of which the Turing instability is an archetype---constitute an important class of dynamical processes occurring in biological, ecological and chemical systems. Recently, it has been shown that the Turing instability can induce pattern formation in discrete media such as complex networks, opening up the intriguing possibility of exploring it as a generative mechanism in a plethora of socioeconomic contexts. Yet, much remains to be understood in terms of the precise connection between network topology and its role in inducing the patterns. Here, we present a general mathematical description of a two-species reaction-diffusion process occurring on different flavors of network topology. The dynamical equations are of the predator-prey class, that while traditionally used to model species population, has also been used to model competition between antagonistic ideas in social systems. We demonstrate that the Turing instability can be induced in any network topology, by tuning the diffusion of the competing species, or by altering network connectivity. The extent to which the emergent patterns reflect topological properties is determined by a complex interplay between the diffusion coefficients and the localization properties of the eigenvectors of the graph Laplacian. We find that networks with large degree fluctuations tend to have stable patterns over the space of initial perturbations, whereas patterns in more homogenous networks are purely stochastic. 
\end{abstract}

\maketitle

\section{Introduction}
Pattern formation is a fundamental natural phenomenon that has abundant examples in biological, ecological and chemical processes \cite{Castets1990,Ouyang1991,Kondo1995,HansMeinhardt2000,Sayama2003,Maini2006}. The first mathematical description was proposed by Alan Turing \cite{Turing1952} where he demonstrated the spontaneous emergence of periodic spatial patterns from a homogeneous equilibrium, driven by the reaction and diffusion of two chemical species corresponding to activators and inhibitors~\cite{Baurmann2007,Rietkerk2008,Harris2005}.  

The existence of pattern formation on discrete media was first introduced by Othmer and Scriven in the context of morphogens diffusing over a network of intercellular connections~\cite{Othmer1971}. Plane-wave driven Turing instability was studied on one and two dimensional lattices, where the wave-functions and wave-numbers in continuous media are replaced by their discrete analogs corresponding to the eigenvalue and eigenvectors of the Laplacian matrix. In recent years, Turing patterns have been shown to exist in reaction-diffusion processes occurring on complex networks, where nodes in the graph are assigned an initial concentration of chemical species, and diffusion occurs along the edges connecting the nodes~\cite{Ide2016,Wolfrum2012}. A small perturbation to the uniform state triggers the growth of Turing patterns above a critical threshold, corresponding to the ratio of the diffusion constants of the respective species. The patterns in this context, correspond to distinct populations of nodes differentiated by their levels of chemical concentrations, and exhibit properties quite different from the classical case. For instance, multiple coexisting stationary states can occur, and hysteresis effects are present, indicating that the patterns are not particularly robust over the space of initial perturbations~\cite{Nakao2010}. 

Turing patterns have also been observed in more exotic variants of network topologies, such as directed, multiplex and time-evolving networks~\cite{Asllani2014,Asllani2014b,Petit2017}. Indeed, given the widespread prevalence of networks across a plethora of socioeconomic, biological and technological systems~\cite{Newman2010,Albert2002,Pastor2004,Cohen2010}, the developed framework is being used to model interesting dynamical phenomena; in \cite{Vidal-Franco2017}, predator-prey \cite{Vanag2009} dynamics is used along with cross-diffusion (in addition to ordinary diffusion), to model the evolution of two competing languages on a scale free network~\cite{Caldarelli2010}. In~\cite{Hata2014} three species interactions are used to model ecological meta-populations where concentrations correspond to the population densities of species in a food web. 

While a lot has been studied regarding the application and existence of Turing patterns in different flavors of large networked systems, there remains much to be understood regarding the precise role of network topology. For instance, it is known that heterogeneities in the connectivity structures (degree-distributions) of networks play a major role in dynamical processes such as percolation, diffusion and epidemic spreading among others~\cite{Callaway2000,Pastor2001,Colizza2007,Dorogovtsev2008,Ghoshal2011}. Although a connection has been established between degree-fluctuations and the localization of the Laplacian eigenvectors that play a role in pattern-formation~\cite{Nakao2010,Hata2017}, the space of topologies in which this has been studied is rather limited. Previous investigation on large networks have typically interpolated between the Erd\H{o}s-R\'{e}nyi (ER) random graph~\cite{Erdos1959} (where fluctuations vanish in the thermodynamic limit) and Barab\'asi-Albert networks that provide a very limited range for probing heterogenous structure~\cite{Barabasi1999}. Furthermore, the properties of the patterns have primarily been studied at the critical threshold (the point at which the Turing pattern emerges) and a rich set of potential dynamical behavior above threshold has not been studied in detail. 

Here, we present an exhaustive analysis of the connection between network topology and the dynamical parameters, uncovering new insights on the nature of Turing patterns in networks.  We find that in certain instances, network topology plays a role not only in instigating the spontaneous differentiation of nodes, but also provides clues to the steady state concentrations. Specifically we show, that while the onset of the pattern can be controlled by tuning the average connectivity of the network, the nature of the final patterns is determined by a complex interplay between degree fluctuations and the diffusion coefficients. In those networks with peaked degree distributions, the patterns are purely stochastic, being determined primarily by perturbations to the initial uniform state. On the other hand, networks with large degree fluctuations induce patterns that are quite robust to initial perturbations---the structure of the network is correlated strongly with the final concentrations of the nodes. This correlation gradually vanishes as a function of increasing the diffusion coefficients such that the patterns are indistinguishable from those formed by homogenous network topologies. We end by discussing the implications of our results in applying reaction-diffusion networks to model real world systems.

\section{Reaction-Diffusion Model and Turing Instabilities on Networks}

Pattern-forming, reaction-diffusion systems, in continuous media, are typically described by a set of two partial differential equations of the form

\begin{eqnarray}
\frac{du(\vec{x},t)}{dt} &= f(u,v) + \epsilon \nabla^2 u(\vec{x}, t)  \nonumber \\
\frac{dv(\vec x, t)}{dt} &= g(u,v) + \gamma \nabla^2 v(\vec{x}, t).
\label{eq:diffc}
\end{eqnarray}
Here $u(\vec{x},t),v(\vec{x},t)$ correspond to the local concentrations of two chemical species, $f(u,v), g(u,v)$ specify their local dynamics and $\epsilon, \gamma$ are the corresponding diffusion coefficients~\cite{Prigogine1968,Yin2013,Lou1998}. Typically $u$ corresponds to an activator, that grows through autocatalytic growth, and $v$ an inhibitor that suppresses $u$. The system is initially considered to be at an uniform steady state ($u_0$, $v_0$)  where $f(u_0,v_0) = g(u_0,v_0) = 0$. The Turing instability spontaneously emerges above a critical threshold of the ratio of diffusion constants, $\sigma= \gamma/\epsilon$ and corresponds to alternating spatial regions of high-  and low-concentrations of $u$.

\begin{figure}[t!]
	\centering
	\includegraphics[width =\columnwidth]{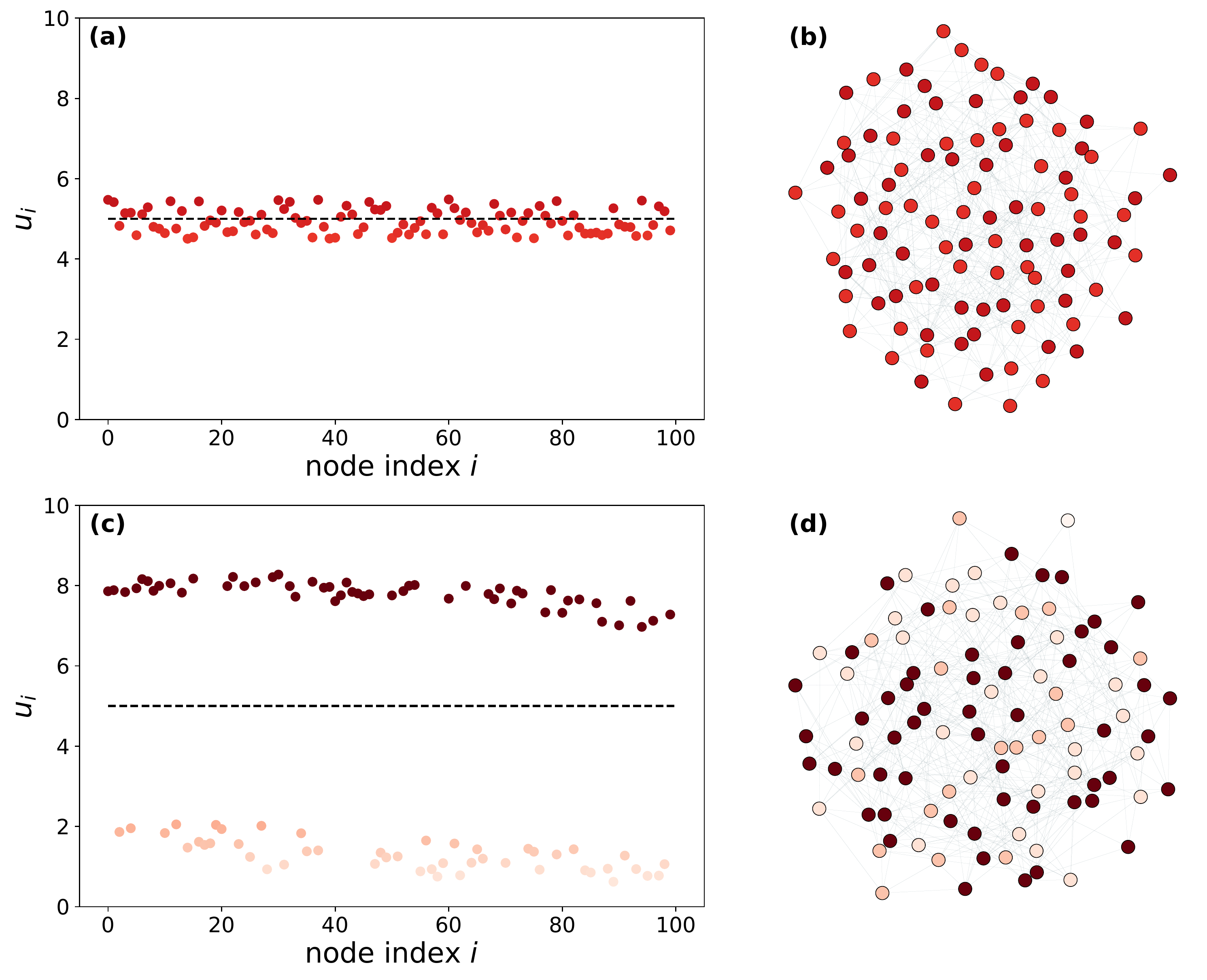}
	\caption{
		Evolution of Turing Patterns on a ER Network with N=100. {\bf(a)} the concentration $u_i$ of node $i$ (sorted in decreasing order of degree $k_i$) immediately after perturbing the uniform background. {\bf (b)} The network with nodes colored according to concentration showing a single population. {\bf(c)} The concentrations after the system has evolved to its steady state, indicating two distinct populations of nodes corresponding to concentrations lying above and below the initial uniform concentration. {\bf(d)} the corresponding network representation.}
\label{fig:turing_demo}		
\end{figure}

A similar set of equations can be used if the system is not continuous but instead is composed of $N$ independent nodes that interact via diffusive transport over $m$ edges~\cite{Ide2016,Mccullen2016,Nakao2010,Vidal-Franco2017}. The analog of the operator $\nabla^2$ is now the Laplacian matrix,

\begin{equation}
	L_{ij} = A_{ij} - k_i \delta_{ij},
	\label{eq:lm}
\end{equation}
where $A_{ij}$ is the symmetric adjacency matrix, whose elements are 1 if there is an edge between nodes $i$ and $j$ and 0 otherwise; $k_i = \sum_j A_{ij}$ is the degree of node $i$ (number of connections) and $\langle k \rangle = 2m/N$ is the average degree of the network. The diffusive transport of chemical species to node $i$ is the sum of all the incoming fluxes from its neighbors $j$ and proportional to their concentration differences. With this modification, Eq.~\eqref{eq:diffc} is transformed into a set of $N$ differential equations ($i = 1 \dots N$) thus,

\begin{eqnarray}
\frac{du_{i}}{dt} &= f(u_i,v_i) + \epsilon \sum_{j=1}^N L_{ij} u_j  \nonumber \\
\frac{dv_{i}}{dt} &= g(u_i,v_i) + \sigma \epsilon \sum_{j=1}^N L_{ij} v_j.
\label{eq:diffn}
\end{eqnarray}
For the purposes of our analysis, we use the Mimura-Murray model which has been used to model predator-prey populations~\cite{Mimura1978}. Our choice is motivated by its ubiquity in studying reaction-diffusion dynamics in networked systems, however our results are broadly applicable to a wider range of dynamics. 

The Turing instability is examined through a linear stability analysis of the uniform stationary state with respect to non-uniform perturbations. While in continuous media, uniform perturbations are decomposed into a set of spatial Fourier
modes representing plane waves with different wave-numbers, in networks their analogs are the eigenvectors $\phi_i ^{(\alpha)}$ and eigenvalues $\Lambda_{\alpha}$ of the Laplacian matrix, where $\alpha = 1,\ldots N$ corresponds to the eigenmode. The eigenvalues $\Lambda_{\alpha}$ are sorted in decreasing order  $\Lambda_{1} >\Lambda_{2} \dots>\Lambda_{N}$ and the first eigenvalue is always zero ($\Lambda_{1} = 0$). Introducing small perturbations $(\delta u_i, \delta v_i)$, substituting into Eq.~\eqref{eq:diffn}, and expanding over the set of the Laplacian eigenvectors, the linear growth rate $\lambda_{\alpha}$ for each node is calculated from a polynomial equation of the form  
\begin{equation}
\lambda_{\alpha}^2 + b(\Lambda_{\alpha})\lambda_{\alpha} + c(\Lambda_{\alpha}) = 0,
\label{eq:charac}
\end{equation}
where $ b(\Lambda_{\alpha}), c(\Lambda_{\alpha})$ are functions of the diffusion coefficients as well as $f, g$. The Turing instability occurs when \emph{at least one of the modes} becomes unstable, indicated by Re $\lambda_{\alpha} >0$ which happens when $c(\Lambda_{\alpha})<0$  (See Appendix~\ref{sec:app} for details of the calculation).  In Fig.~\ref{fig:turing_demo} we show an example of Turing patterns in an ER network of 100 nodes. Panel {\bf a} shows the initial concentration $u_i$ for each node $i$ (sorted in decreasing order of degree) and in {\bf b} we show the network with the nodes colored according to the concentrations. After perturbations, spontaneous differentiation occurs and the system evolves to its final state as shown in panels {\bf c}  and {\bf d}. 

\begin{figure}[t!]
	\centering
	\includegraphics[width =\columnwidth]{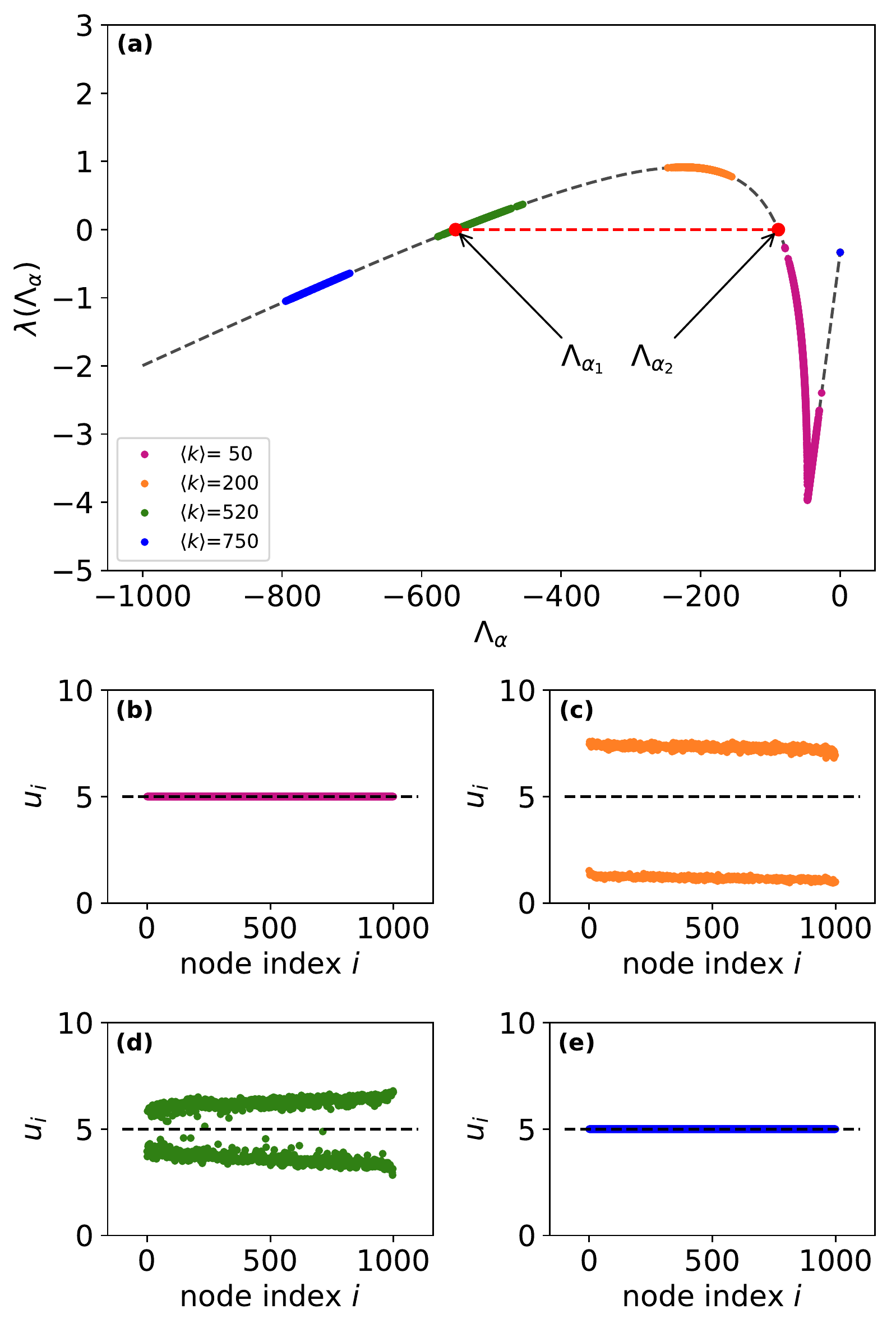}
	\caption{{\bf (a)} Eigenvalue Distributions of four ER networks with $N = 10^3$ and varying average degrees (indicated by color) plotted on the curve $\lambda(\Lambda_{\alpha})$ Eq.~\eqref{eq:instability} (shown as dashed line) corresponding to the Mimura-Murray model with diffusion coefficients $\epsilon = 5\times 10^{-3}$ and $\sigma = 30$. The instability range is marked by $\Lambda_{\alpha_1}, \Lambda_{\alpha_2}$ and is fixed by the diffusion coefficients. The extent of overlap of the growth factors with the instability regime is controlled by $\langle k \rangle$ which shows a non-monotonic trend. {\bf (b)-(e)} the corresponding steady state concentrations for each of the considered networks: $\langle k \rangle= 50, 200, 520, 720$; the black dashed line represents the initial uniform concentrations. Maximum differentiation occurs at $\langle k \rangle= 200$ with an amplitude $A$ = 97.52, which corresponds to the growth-factors being distributed around the peak of the instability curve.}
\label{fig:avg_deg_eig}
\end{figure}

Above threshold, the instability range is parameterized by the roots $\Lambda_{\alpha_1}, \Lambda_{\alpha_2}$ of the parabola $c(\Lambda_{\alpha})$ which takes on negative values in the range $\Lambda_{\alpha_1} \leq c(\Lambda_{\alpha}) \leq \Lambda_{\alpha_2}$ (See Appendix~\ref{sec:app}). The emergence of positive growth rates $\lambda_{\alpha}$ corresponds to the overlap of the Laplacian eigenvalues with the instability regime. Given that $Tr(L) = \sum_i k_i = \sum_{\alpha} \Lambda_{\alpha} $, we have that $\langle k \rangle = \langle \Lambda \rangle$. Thus for a fixed set of diffusion coefficients and dynamical parameters, Turing instabilities can be triggered in \emph{any flavor of network topology} by tuning the average connectivity of the system. Conversely, for fixed network topology, the diffusion coefficients can be tuned (within physically meaningful bounds) to precipitate the instabilities. 

Indeed, it has been pointed out that Turing patterns can be enhanced, as measured by the amplitude of separation $A = [\sum_{i=1}^{N}   {(u_i - u_0)^2 + (v_i - v_0)}]^{1/2}$, by increasing the connectivity of the network~\cite{Vidal-Franco2017}. However the effect is far more nuanced as shown in Fig.~\ref{fig:avg_deg_eig}{{\bf a} where we plot the growth factors as a function of the Laplacian eigenvalues for four different ER networks of varying connectivity. The instability range (delineated by $\Lambda_{\alpha_1}, \Lambda_{\alpha_2}$) is fixed by the diffusion coefficients. For a network of $\langle k \rangle = 50$, we see that none of the growth factors are positive and consequently there is no  change in the initial concentrations (Fig.~\ref{fig:avg_deg_eig}{\bf b}). As the average connectivity increases to 200, we see that all growth factors are positive and spontaneous differentiation occurs, with a large separation between the concentrations (Fig.~\ref{fig:avg_deg_eig}{\bf c}). As the connectivity is increased even further, there is a mixture of positive and negative growth factors, leading to less pronounced differentiation  (Fig.~\ref{fig:avg_deg_eig}{\bf d}). Finally, as connectivity increases further, all growth factors are negative, and no differentiation is apparent (Fig.~\ref{fig:avg_deg_eig}{\bf e}). Note that the maximum of the amplitude $A$ occurs near the peak of the instability curve. Thus, while $\langle k \rangle$ can be tuned to trigger the Turing instability, it is the location of the eigenmodes in the instability regime that determines the trend and amplitude of the differentiation. 

\section{Eigenvector Localization}

Having established the role of the average connectivity in triggering the Turing instability, we next investigate the connection with degree heterogeneities. To do so, it is instructive to outline the role that the eigenmodes and eigenvectors play in the evolution of the system. After perturbing the system, temporal evolution progresses via the change in chemical concentrations  until they reach their steady state. The instantaneous evolution is characterized by  $u_i (t) = u_0 + \delta u_i (t)$, where $\delta u_i (t) = \sum_{\alpha = 1}^{N} a_{\alpha} \exp\left(\lambda_{\alpha} t \right) \phi_{i}^{(\alpha)}$ for some constants $a_{\alpha} $. For negative growth factors, $\lambda_{\alpha}$, the perturbations to the initial state vanish, implying that the evolution of the system can be characterized by considering contributions to the sum for only those modes where $\lambda_{\alpha} >0$.  

Without loss of generality, we set $ a_{\alpha} = 1$ (given the negligible contribution to the sum), such that in the steady state we have

\begin{equation}
  \delta u_i (t_{\textrm{steady}}) \approx \sum_{\lambda_{\alpha} > 0} \exp\left(\lambda_{\alpha} t_{\textrm{steady}}\right) \phi_{i}^{(\alpha)}.
  \label{eq:uapprox}
 \end{equation} 
Thus, differentiation of nodes is associated with the properties of the eigenvectors corresponding to positive growth factors. The extent to which the eigenvectors are connected to network topology depends on its localization properties~\cite{McGraw2008,Pastor-Satorras2016,Jalan2011}. A localized vector has the majority of its normalization weight concentrated on a small subset of its components, whereas a delocalized vector has its weights distributed relatively uniformly across all of its components. An ideal way to quantify the localization of a vector $\textbf{V}$ of size $N$, is to measure its Inverse Participation ratio (IPR) defined as 
\begin{equation}
P(\textbf{V}) =\frac{ \sum_{i=1}^{N} V_i^4}{(\sum_{i=1}^{N} V_i^2)^2}.
\label{eq:ipr}
\end{equation}
The IPR lies in the range $1/N \leq P\left(\textbf{V}\right) \leq 1$ with the upper limit corresponding to all weights localized on a single component, whereas the lower limit represents the situation where the weights are uniformly distributed over all components.

\begin{figure}[t!]
\raggedleft
\includegraphics[width = \columnwidth]{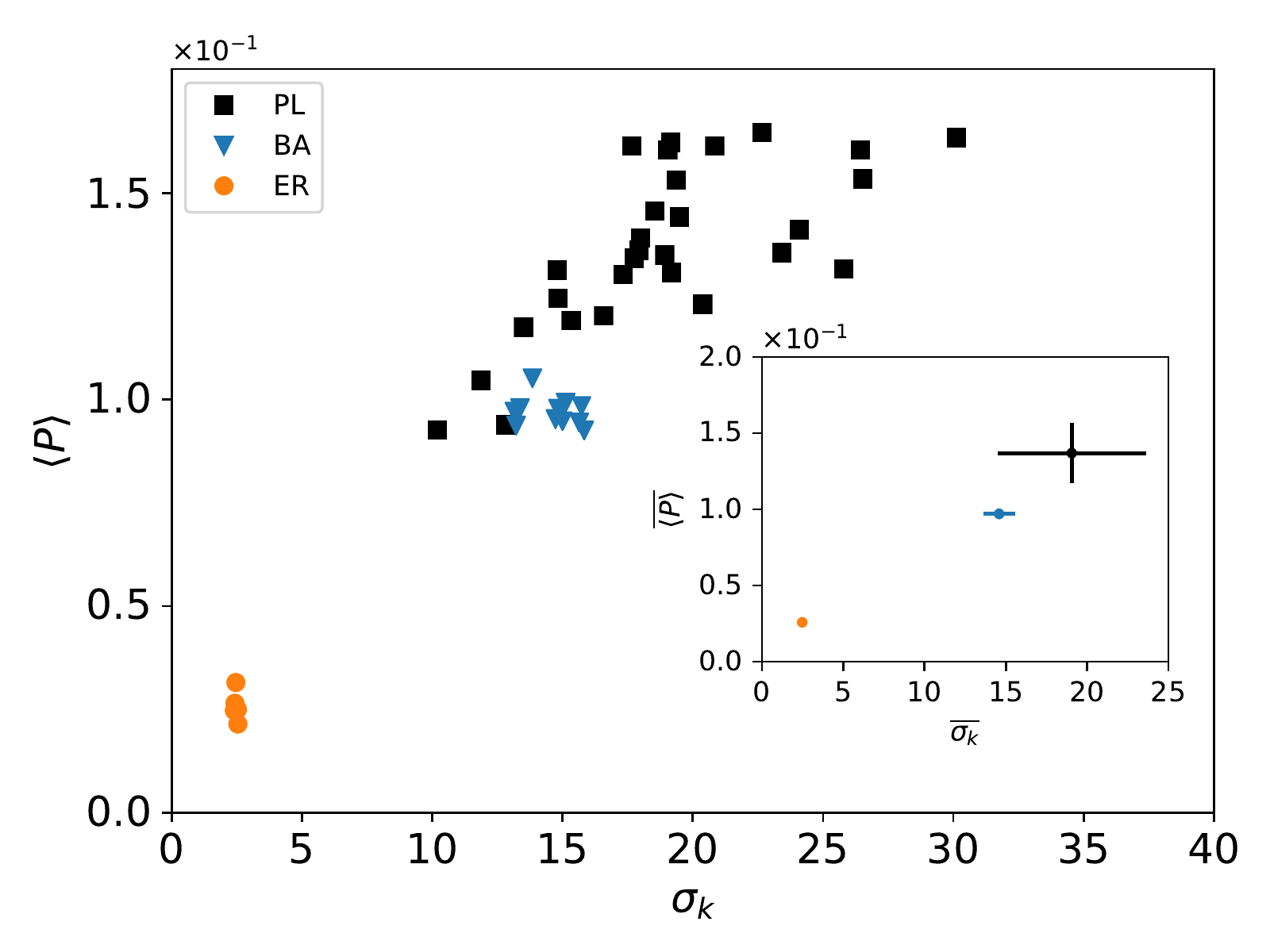}
\caption{Network topology and eigenvector localization. The average IPR $\langle P \rangle$ (Eq.~\eqref{eq:avipr}) as a function of degree fluctuations $ \sigma_k =\sqrt{ \langle k^2 \rangle - \langle k \rangle^2} $ for BA,ER and PL networks with $N=10^3$ and $\langle k \rangle \approx 6.0$. While the ER network has a binomial degree distribution, the BA network goes as $p_k \sim k^{-3}$ in its tail and the PL network has a distribution of the form $p_k = k^{-\beta}/\zeta(\beta, x_{min})$ where $\zeta$ is the Hurwitz zeta function and $\beta = 2.1, x_{min} = 2$. The inset shows the average of data points indicating a clear monotonic trend of the eigenvector localization with increasing degree heterogeneity.}
\label{fig:iprsigma}
\end{figure}

There is a clear correlation between node degree and the Laplacian eigenvalues in networks~\cite{Hata2017}. In addition, eigenvectors associated with higher degree nodes are strongly localized, whereas the localization is weaker for nodes with fewer connections. To see the effects clearly in terms of network topology, we generated a series of networks using the BA model, the ER random graph and power law (PL) networks ($p_k \sim k^{-\beta}$) using the configuration model~\cite{Molloy1995,Molloy1998}. The networks are created with different random seeds and the relevant parameters are tuned such that they all have the same average degree $\langle k \rangle$ but differ significantly in their degree fluctuations $\sigma_k =\sqrt{ \langle k^2 \rangle - \langle k \rangle^2}$. For each generated network, we compute the average IPR 
\begin{equation}
\langle P \rangle=  \frac{1}{N} \sum_{\alpha = 1}^{N} P(\textbf{V}_{\alpha}),
\label{eq:avipr}
\end{equation}
and plot it as a function of $\sigma_k$ as shown in Fig.~\ref{fig:iprsigma}. A clear monotonic trend is seen, where the PL networks with large degree fluctuations have significantly higher $\langle P \rangle$ as compared to the ER networks. Taking the average of all points (shown as inset) shows a  marked difference between the BA networks ($p_k \sim k^{-3}$) and the PL graphs ($p_k \sim k^{-2.1}$), given that $\sigma_k \rightarrow \infty$ for $\beta < 3$ in the thermodynamic limit.  Thus increasing degree heterogeneities leads to increasing localization of the Laplacian eigenvectors of the network. 

Earlier work~\cite{Nakao2010,Hata2014} connect the eigenvector $\phi_i ^{(\alpha_c)}$ associated with the critical eigenmode $\alpha_c$(corresponding to the onset of the Turing instability) with the differentiation patterns in early phases of the evolution, with the steady-state concentrations $\delta u_i (t_{\textrm{steady}})$ and $\delta v_i (t_{\textrm{steady}})$, being determined by non-linear effects. However, as we show next, if contributions to Eq.~\eqref{eq:uapprox} are composed of localized eigenvectors, then a connection can be made to the resultant Turing Patterns and topological properties of the network (quantified by $\sigma_k$). If on the other hand, most of the sum is determined by non-localized vectors then the effect of the network topology is washed out and patterns emerge randomly, primarily as  function of initial perturbations to the uniform background.

\section{Numerical Results}

Next, we investigate the typology of the emergent Turing patterns as a function of degree heterogeneity. We start with the limiting cases of the star- and complete-graphs and fill the spectrum, interpolating between the ER and PL networks.

\begin{figure}[b!]
	\centering
	\includegraphics[width=\columnwidth]{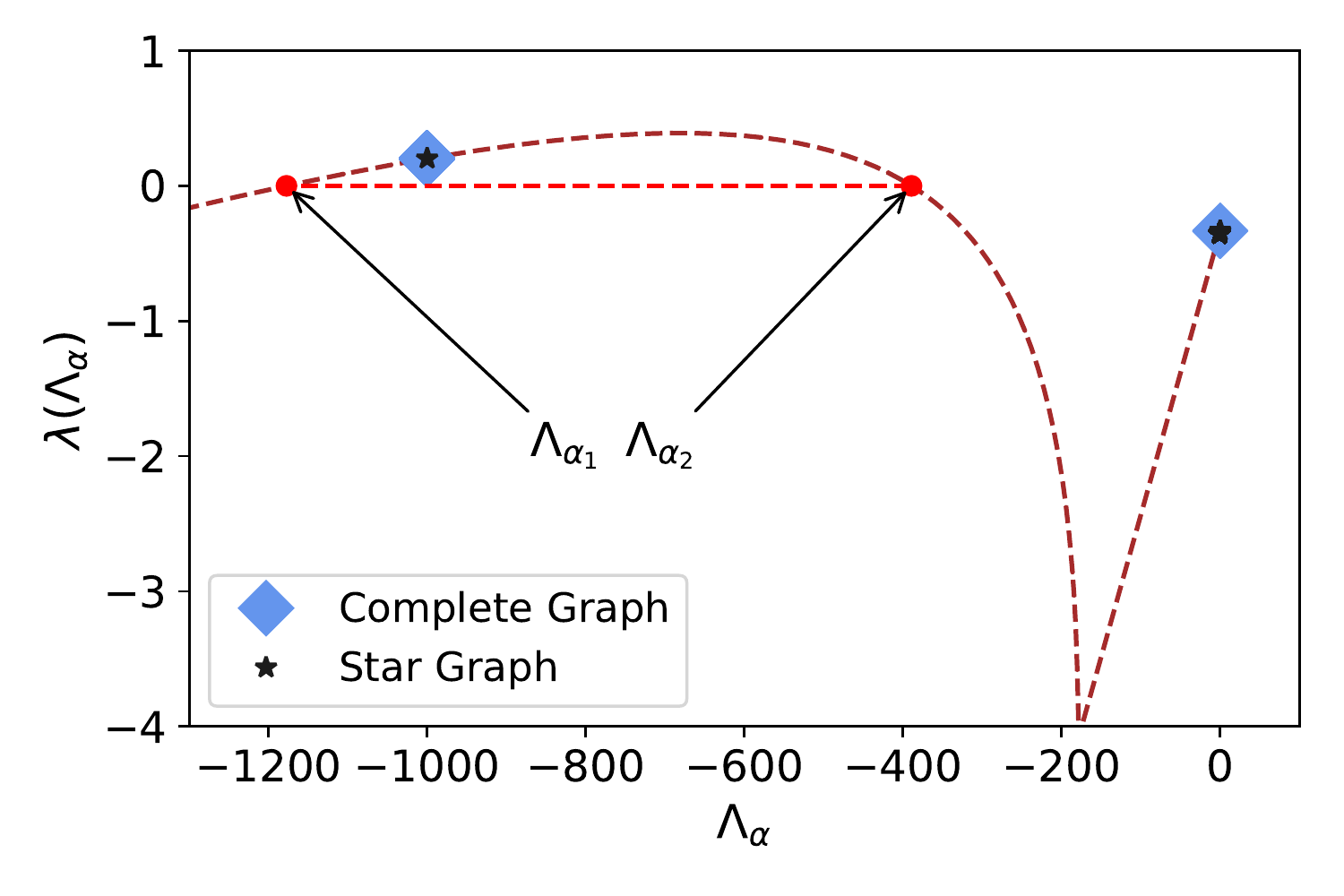}
	\caption{Eigenvalue distributions of the star and complete graphs with $N=10^3$, plotted on the curve $\lambda(\Lambda_{\alpha})$  Eq.~\eqref{eq:instability} (shown as dashed line) corresponding to the Mimura-Murray model with diffusion coefficients $\epsilon = 2\times 10^{-3}$ and $\sigma = 20$. The instability range is marked by $\Lambda_{\alpha_1}, \Lambda_{\alpha_2}$. The star graph has only 1 positive growth factor whereas the complete graph has 999 positive overlapping growth factors.}
\label{fig:eigstarcomp}	
\end{figure}

\subsection{Limiting cases}
A star graph consists of $N$ nodes with a central node connected to $N-1$ peripheral nodes, with no additional connection between them. The average degree $\langle k \rangle =  2\left(1  - 1/N\right) $ and the second moment is $\langle k^2 \rangle =  N  - 1$. Consequently the fluctuations are extensive and of the form $\sigma_k^{\textrm{star}} \approx \sqrt{N}$. On the other end of the spectrum is the complete graph where all pairs of nodes are connected to each other. In this case $\langle k \rangle = N-1$ and $\langle k^2 \rangle =  (N  - 1)^2 $ and therefore $\sigma_k^{\textrm{complete}} = 0$. 

 \begin{figure}[t!]
	\centering
	\includegraphics[width=\columnwidth]{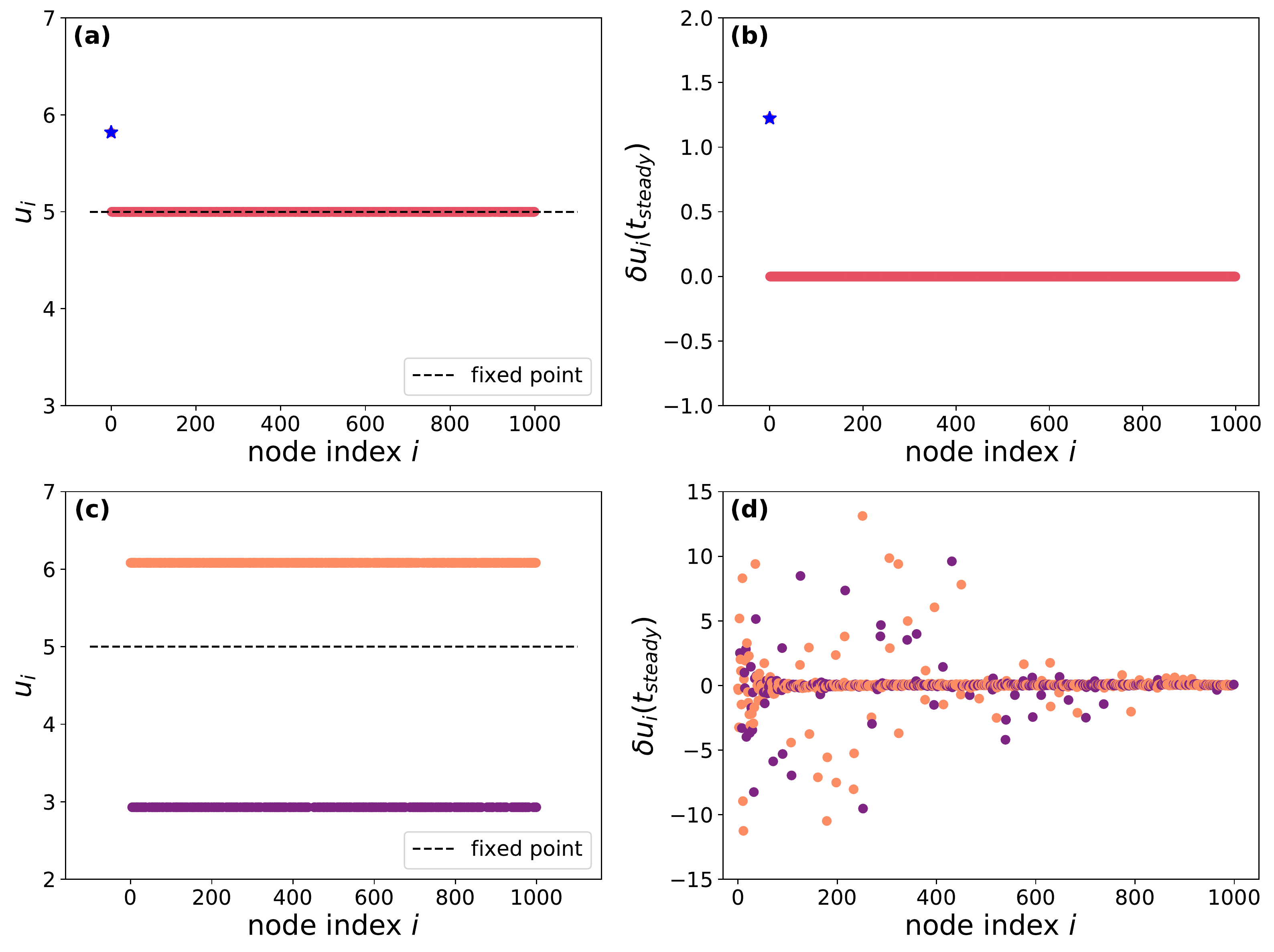}
	\caption{Steady state chemical concentrations $u_i$ for each node $i$ in the star graph \textbf{(a)} and complete graph \textbf{(c)}. Nodes are arranged in decreasing order of degree and the dynamical parameters are the same as Fig.~\ref{fig:eigstarcomp}. The dashed black line represents the initial concentrations. \textbf{(b),(d)} The contributions of the eigenvectors associated with positive growth factors (Eq.~\eqref{eq:uapprox}) for each of the networks. Nodes are colored according to their final concentration values.  The final Turing pattern for the star graph is determined exclusively by the highly localized  leading eigenvector corresponding to the highest degree node $\left(\langle P_{\lambda_{\alpha^{+}}}\rangle = 1\right)$. There appears little correlation between the graph topology and the final Turing pattern in the complete graph, with nodes differentiating randomly $\left(\langle P_{\lambda_{\alpha^{+}}}\rangle = 4.7 \times 10^{-2}\right)$.}
\label{fig:starcomp}	
\end{figure}

We simulate Mimura-Murray dynamics on each of the networks with diffusion constants $\epsilon = 2\times 10^{-3}$ and  $\sigma = 20 $ to induce differentiation via the Turing instability.  Plotting the growth factors as a function of the Laplacian eigenvalues (Fig.~\ref{fig:eigstarcomp}) reveals that the star graph has one positive growth factor, whereas the complete graph has $N-1$ positive growth factors. This is reflected in the final concentrations for each of the networks as shown in Fig. ~\ref{fig:starcomp}{\bf a,c}. In the star graph the only node that differentiates is the central node, whereas in the complete graph two separate clusters of roughly equal size emerge.  Approximating the final concentrations using Eq.~\eqref{eq:uapprox} we see that in Fig.~\ref{fig:starcomp}{\bf b}, the Turing pattern formed in the star graph is described entirely by the eigenvector corresponding to the central node with the maximum number of connections.  Conversely, as seen in Fig.~\ref{fig:starcomp}{\bf d}, there appears to be no connection with the final Turing pattern and the structure of the eigenvectors associated with differentiation. Nodes differentiate randomly as a function of the initial perturbations. This sharply contrasting behavior is strongly connected to the varying localization properties in each flavor of network. Calculating the average IPR of the eigenvectors associated with positive growth factors $\langle P_{\lambda_{\alpha^{+}}}\rangle$, we see that the star graph has a perfectly localized eigenvector $\langle P_{\lambda_{\alpha^{+}}}\rangle = 1$, whereas the complete graph has $\langle P_{\lambda_{\alpha^{+}}}\rangle = 4.7 \times 10^{-2}$. 
 
\subsection{Dynamics on ER and PL Networks}
The cases considered thus far constitute the two end points of the spectrum of network structure in terms of their degree heterogeneities. For maximal degree fluctuations, the Turing pattern is determined entirely by topological effects, whereas for no degree fluctuations, the patterns are random and have little-to-no connection with network topology. To determine the extent to which topological effects manifest themselves, we next interpolate between these two limits, starting with the ER network and then two power law networks generated via the configuration model; one with non-extensive fluctuations $\left (p_k \sim k^{-3.1}\right)$, and the other with extensive fluctuations $(p_k \sim k^{-2.1})$.  The relevant parameters are tuned such that all networks have the same average degree $\langle k \rangle \approx 7$.
 
\begin{figure}[b!]
\centering
\includegraphics[width=\columnwidth]{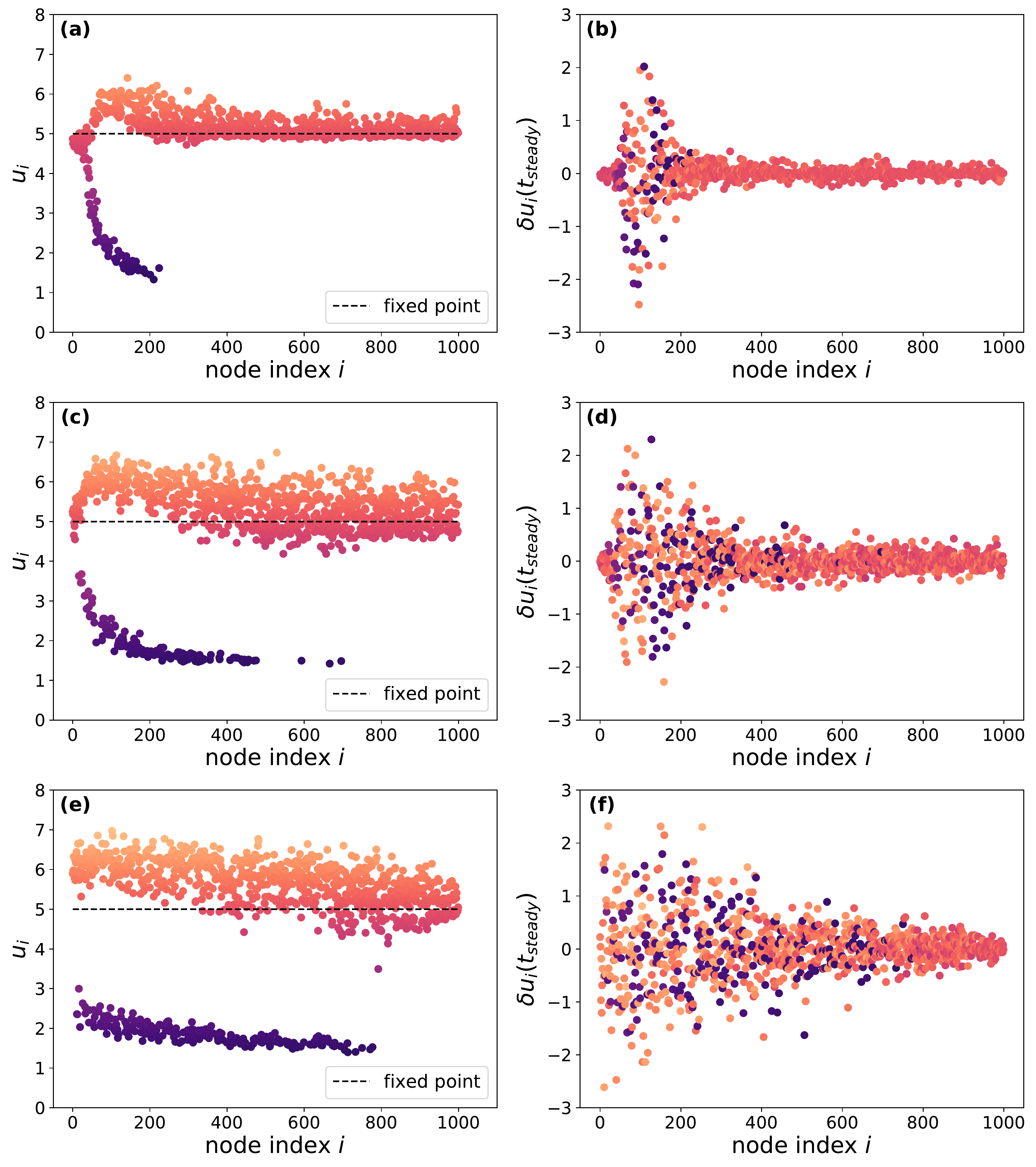}
\caption{Steady state chemical concentrations $u_i$ for each node $i$ in the power law graph with parameters  $\beta = 2.1, x_{min} = 2$ \textbf{(a)}, power law graph with parameters  $\beta = 3.1, x_{min} = 4$ \textbf{(c)} and ER network \textbf{(e)}. The diffusion coefficients are $\epsilon = 1.1 \times 10^{-1}, \sigma = 17$. All networks have $\langle k \rangle \approx 7$ and nodes are arranged in decreasing order of degree. The dashed black line represents the initial concentrations. \textbf{(b),(d),(e)} The contributions of the eigenvectors associated with positive growth factors (Eq.~\eqref{eq:uapprox}) for each of the networks. Nodes are colored according to their final concentration values. The localization of the eigenvectors associated with positive growth factors $\langle P_{\lambda_{\alpha^{+}}}\rangle$ are from (top to bottom) $2 \times 10^{-1}, 6.1 \times 10^{-2},1.7 \times 10^{-2}$.}
\label{fig:netcomp}
\end{figure}

\begin{figure*}[t!]
\centering 
\includegraphics[width=0.7\textwidth]{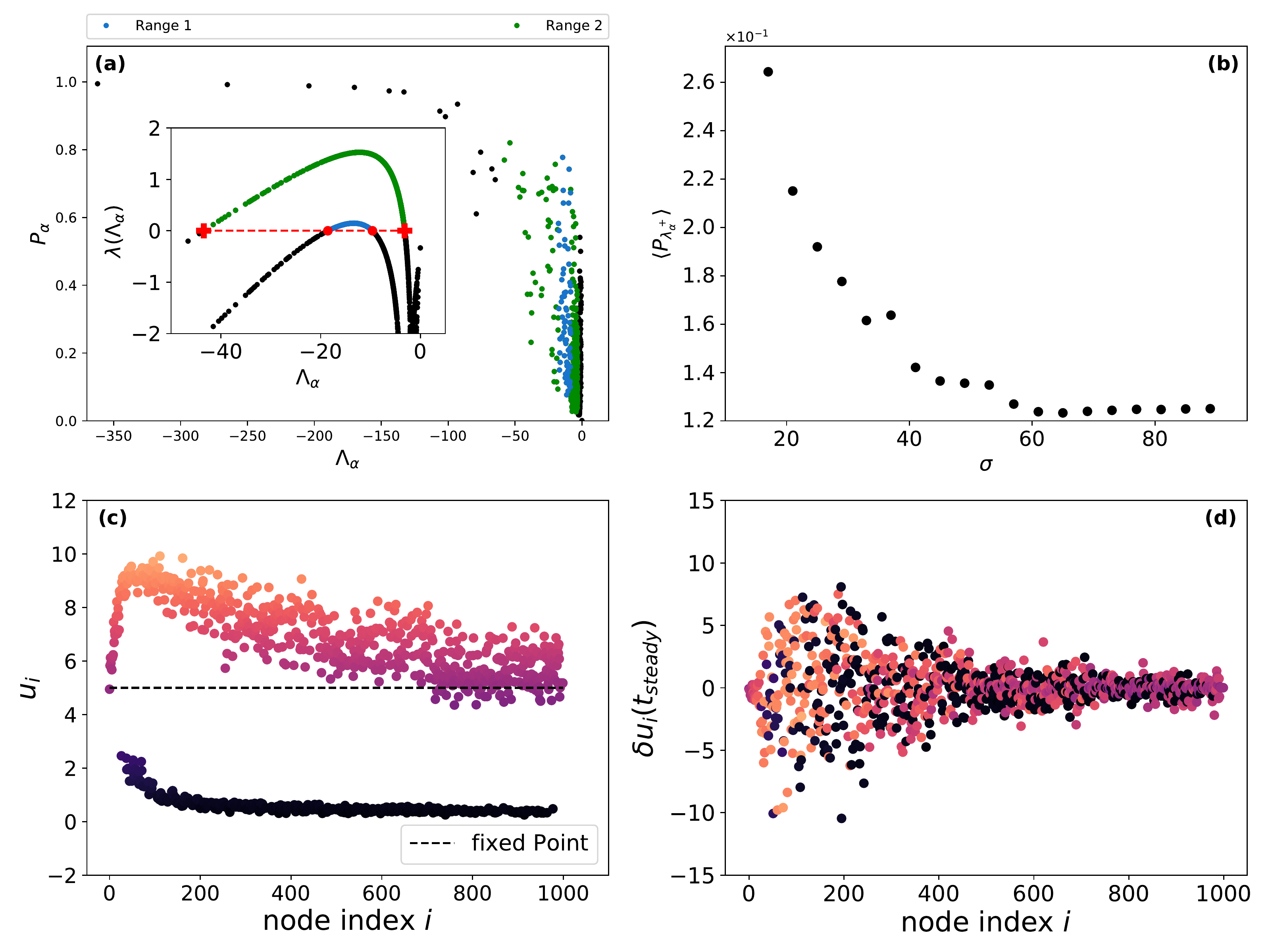}
\caption{Effect of diffusion on Turing patterns. {\bf (a)} The IPR (Eq.~\eqref{eq:ipr}) as a function of the Laplacian eigenvalues of a PL network with $\beta = 2.1, x_{min} = 2$, for different values of the diffusion coefficient: \textit{Range 1} corresponds to $\epsilon= 1.1\times 10^{-1}, \sigma = 17$ and \textit{Range 2} corresponds to $\epsilon= 7\times 10^{-2}, \sigma = 55$. Positive growth factors (shown as inset) are colored blue and green. The localization decreases from \textit{Range 1} $\left(\langle P_{\lambda_{\alpha^{+}}}\rangle = 2.2 \times 10^{-1}\right)$ to \textit{Range 2} $\left(\langle P_{\lambda_{\alpha^{+}}}\rangle = 1.6 \times 10^{-1}\right)$.  {\bf (b)} The localization of the unstable modes plotted as a function of $\sigma$ showing a clear delocalization trend (and therefore more stochasticity in pattern formation). {\bf (c)} The steady state concentrations of the nodes corresponding to the wider instability range, showing a markedly different pattern than seen in Fig.~\ref{fig:netcomp}{\bf a}. {\bf (d)} the contributions from the eigenvectors associated with the positive growth factor. Nodes colored according to concentration.}
\label{fig:diffipr}
\end{figure*}

In Fig.~\ref{fig:netcomp} we show the results of our simulations for $\epsilon= 1.1\times 10^{-1}$, $\sigma = 17$ for all three network topologies. Given that both the instability range and the average degree is fixed, the observed variations are purely a function of $\langle k^2 \rangle$ and the attendant localization properties.    For the PL network with extensive fluctuations, the pattern is akin to that seen for the star graph in the sense that there is a close connection between the final concentrations (Fig.~\ref{fig:netcomp}{\bf a}) and the contributions from the eigenvectors associated with node differentiation (Fig.~\ref{fig:netcomp}{\bf b}), although it is not as pronounced as in the star graph, given the comparatively lower localization $\left(\langle P_{\lambda_{\alpha^{+}}}\rangle = 2 \times 10^{-1} \right)$. This correspondence gradually vanishes as fluctuations dissipate and the localization of the eigenvectors decrease $\left(\langle P_{\lambda_{\alpha^{+}}}\rangle = 6.1 \times 10^{-2} \right)$, as seen for the PL network with finite fluctuations (Figs.~\ref{fig:netcomp}{\bf c,d}). Eventually, as we get to the ER network, the behavior is much like that for the complete graph (Figs.~\ref{fig:netcomp}{\bf c,d}), whereby the final concentrations appear uncorrelated with the properties of the relevant eigenvectors $\left(\langle P_{\lambda_{\alpha^{+}}}\rangle = 6.1 \times 10^{-2} \right)$, and are more a function of the random perturbations to the initial state.

\subsection{Interplay between diffusion and topology}

Next, we investigate the role played by the diffusion constants in the generated patterns. Our strategy thus far has been to keep the dynamical parameters fixed ---therefore fixing the instability range $\left(\Lambda_{\alpha_1}, \Lambda_{\alpha_2}\right)$---while probing for the effects of network topology.  Now we fix the network topology (and the corresponding eigenspectrum) and tune the diffusion parameters such that we can change the number of positive growth factors that lead to differentiation. In Fig.~\ref{fig:diffipr}{\bf a}, we plot the IPR of all the eigenmodes (Eq.~\eqref{eq:ipr}) as a function of the eigenvalues of the power law network with degree distribution $p_k \sim k^{-2.1}$. The eigenvalues associated with positive growth factors corresponding to the parameters used in Fig.~\ref{fig:netcomp} are colored blue. Changing the values of the diffusion constants to $\epsilon = 7\times 10^{-2}$ and $\sigma = 55$ leads to a wider instability regime, and the corresponding unstable eigenvalues are colored green. The distribution of positive growth factors for both instances is shown as inset.  

\begin{figure*}[t!]
\centering
\includegraphics[width=\textwidth]{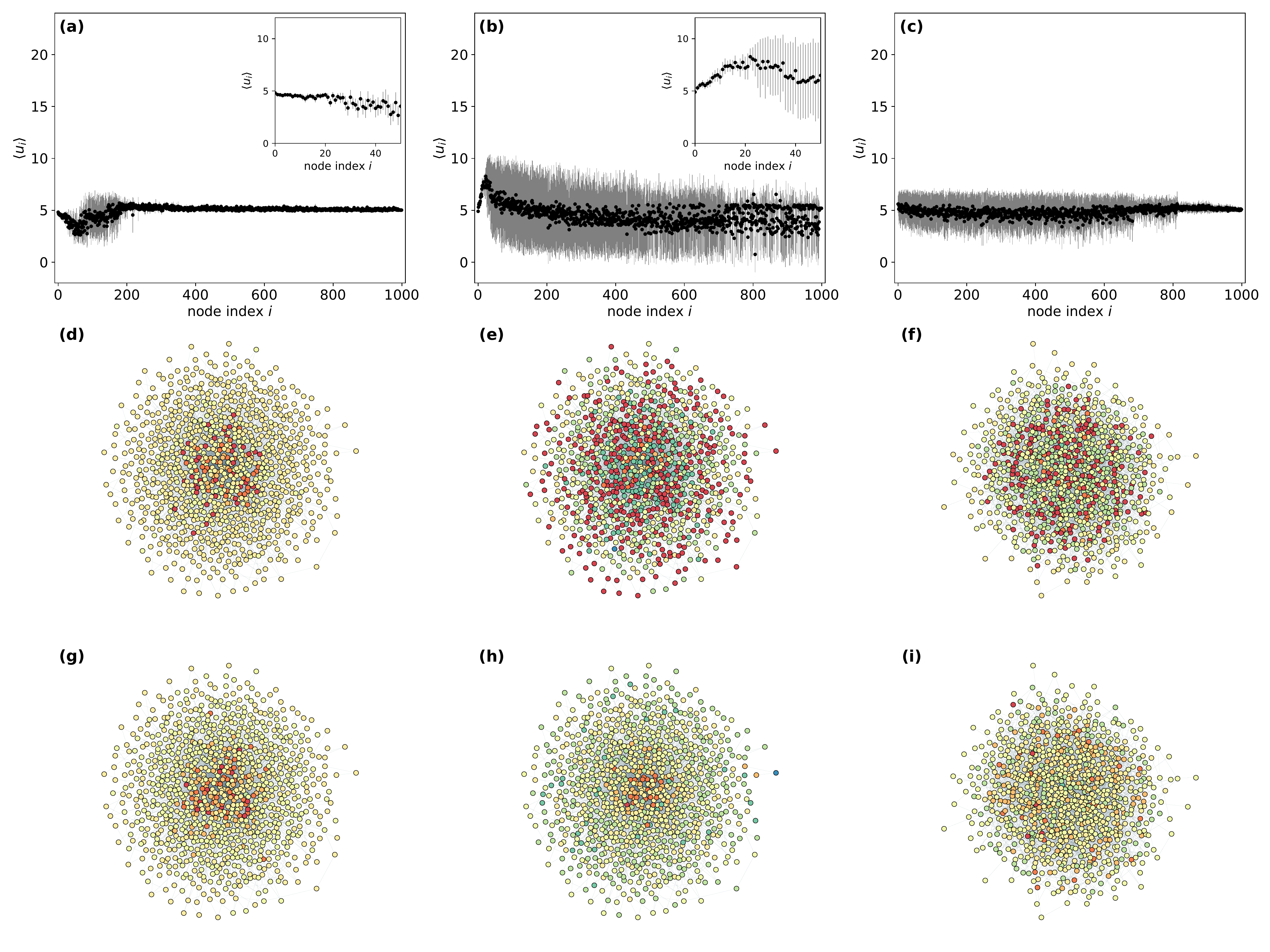}
\caption{Averaged steady state concentrations $\langle u_i\rangle$ Eq.~\eqref{eq:avgu} over multiple realizations of the dynamics: {\bf(a)} The PL network from Fig.~\ref{fig:diffipr} for \textit{Range 1}. {\bf(b)} the same network over \textit{Range 2}. {\bf (c)} The ER network with the same parameters corresponding to \textit{Range 1}. Node indices are sorted in decreasing order of degree, and a blowup of the first 50 nodes is shown as inset. {\bf (d)-(f)} The steady state concentrations for a single realization shown at the network level (ordered the same way as the panel above). Nodes are colored by concentration level. {\bf (g)-(i)} The steady state concentrations, now averaged over a 100 realizations. We see that the differentiation pattern in the PL network for \textit{Range 1} is stable over the space of initial perturbations, whereas the patterns get washed out, either when $\sigma$ is increased, or when degree heterogeneities vanish as in the ER network.} 
\label{fig:netavg}
\end{figure*}

Even though the networks are identical, one sees a larger set of unstable eigenvalues associated with delocalized eigenvectors in the case of the wider instability range. The IPR decreases from $2.2 \times 10^{-1}$ to $1.6 \times 10^{-2}$ and its effect is reflected in the final concentrations shown in Fig.~\ref{fig:diffipr}{\textbf c} which are markedly different than seen in Fig.~\ref{fig:netcomp}{\bf a}. Furthermore, the differentiation patterns are more irregular and are comparatively uncorrelated with the structure of the associated eigenvectors as seen in Fig.~\ref{fig:diffipr}{\bf d}. Indeed, the trend is more akin to seen that for the ER network, where the resultant patterns are purely stochastic and dependent on the initial perturbations to the system. Plotting the localization as a function of increasing $\sigma = \gamma/\epsilon$ (which leads to an increase in the instability range), shows a clear montonically decreasing trend of the eigenvector localization. Thus even in networks where there is a connection between the topological structure and the resultant differentiation patterns, this effect gets washed out depending on the diffusion coefficients. In the specific case of the Mimura-Murray model, this happens when the inhibitor diffuses much faster than the activator. 

\subsection{Averaged patterns}
A clear way to see the effect of topology on the final Turing pattern is to generate multiple realizations of the dynamical process as a function of initial perturbations to the uniform background.  To probe for this, we generated different realizations of the dynamics for the network shown in Fig.~\ref{fig:diffipr}, for the two different instability ranges, as well as the ER network shown in Fig.~\ref{fig:netcomp} and took the average of the resultant concentrations for each node. That is for each node $i$ in the relevant network, we computed 
\begin{equation}
\langle u_i \rangle = \frac{1}{L} \sum_{l=1}^{L} u_i ^{l},
\label{eq:avgu}
\end{equation}
where $l$ corresponds to a single instance of random perturbations, and $L =100$ is the total number of realizations of this process. In Fig.~\ref{fig:netavg} we plot the results of our analysis. Panel {\bf a} shows $\langle u_i \rangle$ for the PL network in the lower instability range. While some fluctuations are apparent, these are associated with the relatively narrow set of differentiating nodes whose concentrations are well approximated with their corresponding eigenvectors. The Turing pattern for the network in a single realization (Fig.~\ref{fig:netavg}{\bf d}) is almost identical to the averaged concentration over a 100 different realizations (Fig.~\ref{fig:netavg}{\bf g}). The degree heterogeneity in this case leads to an element of determinism in the final Turing pattern, that is robust to variations in the perturbations introduced to the initial state. Increasing the diffusion of the inhibitor relative to the activator for the same network, results in a very different situation as seen in Fig.~\ref{fig:netavg}{\bf e}. Fluctuations in the average concentration dramatically increase for practically all nodes, indicating little correlation between network topology and the resultant pattern. This is reflected when comparing the Turing pattern for one realization (Fig.~\ref{fig:netavg}{\bf e}) to the average over multiple realizations (Fig.~\ref{fig:netavg}{\bf h}).The average of the concentrations bear no resemblance to any one instance, indicating that the  patterns are purely stochastic. A similar trend is seen for the ER network with the same diffusion coefficients as the first PL network. Due to negligible degree heterogeneities, the topology has little-to-no effect on the Turing patterns (Fig.~\ref{fig:netavg}{\bf c,f,i}).

\section{Discussion}

Taken together, our results shed new insight on the influence of network topology in the instigation of Turing patterns and the eventual differentiation of nodes in the steady state. The conditions for instability are set by the diffusion coefficients, and the instability emerges when the eigen-spectrum of the network overlaps with the instability regime. In principle node differentiation can be induced in any flavor of network topology, by tuning the average connectivity of the system, which controls the extent of overlap between the eigen-spectrum and the instability regime; conversely for a fixed network topology, one can tune the dynamical parameters---within meaningful bounds---to once again generate Turing patterns. The nature of the eventual pattern is connected deeply with the localization properties of the eigenvectors of the associated network. Networks with large degree fluctuations have highly localized eigenvectors, which for certain regimes of the diffusion parameters correlate strongly with the steady state Turing pattern. Indeed, in the case of the star graph which has maximal degree fluctuations, the differentiation is determined entirely by the maximally localized eigenvector, associated with the central node. On the other end of the spectrum, pattern formation in networks with little-to-no degree fluctuations is entirely stochastic and essentially independent of the specific properties of the eigenvectors of the network. 

Interpolating between heavy-tailed distributions with extensive fluctuations and peaked distributions with non-extensive fluctuations, we see a gradual decoupling between the emergent Turing pattern and the topological properties of the network, with the former being robust to the space of initial perturbations to the uniform background, whereas the latter being entirely influenced by it. Even for those networks with strongly localized eigenvectors, we find that its influence in pattern formation is diminished by tuning the diffusion coefficients. For the specific case of the Mimura-Murray model, the link between topology and pattern formation vanishes either when degree fluctuations are suppressed, or when the inhibitor diffuses at a much faster rate than the activator. In both cases, the extent to which topology matters, can be quantified by calculating the inverse proportionality ratio (IPR) of eigenvectors associated with unstable eigenvalues. 

Our results have interesting implications for the application of Turing patterns in networks, in service of modeling real world dynamical processes. One of the key prevalent phenomenon in recent times is the polarization of opinions, and socioeconomic segregation across many social systems. In the case of opinion formation, one possibility, would be to model the evolution of mutualistic-antagonistic dynamics on social networks, which is similar in flavor to the predator-prey model. It has been suggested that the formation of echo chambers and polarized opinions, is a function of both the speed of information spread (diffusion) and the increased connectivity afforded by the prevalence of social media~\cite{Barbera2015,Del-Vicario2016}. This is indeed borne out by Fig.~\ref{fig:avg_deg_eig}, where we see an increase in the amplitude of the differentiation as connectivity in the network is increased (for that particular combination of diffusion parameters). However, perhaps encouragingly, the differentiation vanishes as the network is even more connected. To the extent that opinion formation can be modeled by the framework presented here, tuning network connectivity appears to be a proscriptive tool, in terms of preventing segregation. 

Furthermore, beyond the question of whether Turing patterns exist on networks, it is worth investigating the extent to which the differentiation properties are associated with specific nodes. In other words, given a set of dynamical parameters, the extent to which a given node takes on a characteristic concentration (opinion, socioeconomic state, language adoption etc.) in the final state. As Fig.~\ref{fig:netavg} indicates, for highly heterogenous network topologies and small diffusion rates, the network has a stable hierarchical structure in terms of the final concentrations. The situation persists for the case when the mutualistic opinion diffuses much faster than the antagonistic one. As the rate of antagonistic opinions spread faster relative to mutualistic ones, the hierarchical structure is lost and while multiple opinions form, they are stochastic with respect to the population on which the information is spreading. Of course, similar considerations apply when modeling ecological species~\cite{Hata2014} or linguistic adoption~\cite{Vidal-Franco2017}.  

\begin{acknowledgments}
SM, GG and JP acknowledge support from the Global Research Network program through the Ministry of Education of the Republic of Korea and the National Research Foundation of Korea (NRF-2016S1A2A2911945). APM and MMJ are supported by the Spanish Ministerio de Econom\'ia y Competitividad and European Regional Development Fund under contract MAT2015-71119-R AEI/FEDER, UE, and by Xunta de Galicia under Research Grant No. GPC2015/014. GG and SM also thank the University of Rochester for financial support. 
\end{acknowledgments}
\bibliography{ref_v2}
\bibliographystyle{apsrev4-1}


\vspace{10px}
\appendix

\appendix

\section{ Dynamical Model  and Linear Stability Analysis }
\label{sec:app}

\setcounter{equation}{0}
\renewcommand{\theequation}{A\arabic{equation}}

The general form of the differential equations used here are,
\begin{equation*}
\frac{du_{i}}{dt} = f(u_i,v_i) + D_{act} \sum_{j=1}^N L_{ij}u_j
\end{equation*}
\begin{equation*}
\frac{dv_{i}}{dt} = g(u_i,v_i) + D_{inh}\sum_{j=1}^N L_{ij} v_j,
\end{equation*}
where $D_{act} = \epsilon$ and $ D_{inh} = \gamma$, with $\gamma = \sigma \epsilon$. For the Mimura-Murray model 
\begin{equation}
f(u,v) = \left(\frac{a +bu -u^2}{c} - v\right)u; \quad g(u,v) = \left(u - (1 + dv)\right)v.
\end{equation}
 Setting the constants $a =35 ,b =16 ,c =9 ,d =2/5 $ yields a fixed point concentration $(u_0,v_0) = (5,10)$. The equations are evaluated using the $4^{th}$ order Runge-Kutta Method. Initial non-uniform perturbations to node concentrations are $\delta u(0) = u_0 + 0.1u_0 r$ and $\delta v(0) = v_0 + 0.1v_0r$ where $r$ is  a uniform random number in the range $-1<r<1$.

The equations can be cast in matrix form thus,

\begin{widetext}
\begin{equation}
\begin{pmatrix} 
\frac{du_i}{dt}    \\ 
\frac{dv_i}{dt}  
\end{pmatrix} =  J|_{u_0,v_0} 
\begin{pmatrix} 
u_i   \\ 
v_i  
\end{pmatrix} + \sum_{j=1}^{N} L_{ij}
\begin{pmatrix} 
D_{act} & 0  \\ 
  0 & D_{inh} 
\end{pmatrix} \begin{pmatrix} 
u_i   \\ 
v_i  
\end{pmatrix} 
\label{eq:umatrix}
\end{equation}
\end{widetext}

where 
\begin{equation*}
J|_{u_0,v_0} =  \begin{pmatrix} 
f_u & f_v    \\ 
g_u & g_v  
\end{pmatrix},
\end{equation*} is the Jacobian matrix and
\begin{equation*}
\textbf{D} = D|_{u_0,v_0} =  \begin{pmatrix} 
D_{act}  & 0  \\ 
0 &  D_{inh}
\end{pmatrix} = \begin{pmatrix} 
\epsilon & 0   \\ 
0 &  \gamma, 
\end{pmatrix}
\end{equation*} is the diffusion matrix both evaluated at the fixed point $(u_0,v_0)$.

\begin{figure*}[t!]
\centering
\includegraphics[width=\textwidth]{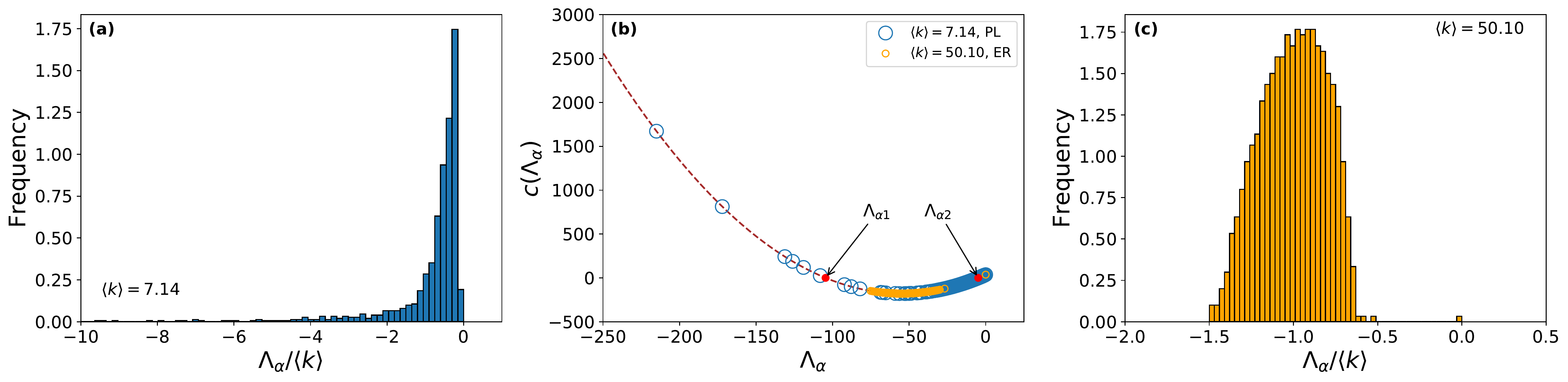}
\caption{Mimura-Murray model for diffusion with constants $\epsilon = 3\times 10^{-2}$, $\sigma = 80$ on ER and Power-Law Networks ($\beta = 2.1. x_{min} = 2$) with $N=10^{3}$. \textbf{(a)} The eigenvalue distribution of the PL and \textbf{(c)} ER networks. The eigenvalues are scaled with respect to the average degree $\langle k \rangle$. \textbf{(b)} The eigenvalues of the ER and PL networks distributed on $c(\Lambda_{\alpha})$ (Eq.~\eqref{eq:calpha}).}
\label{fig:criticality}
\end{figure*}

The corresponding eigenvalue equation for the Laplacian matrix is:
\begin{equation*}
\sum_{i=1}^{N} L_{ij} \phi_{j}^{(\alpha)} = \Lambda_{\alpha}\phi_{j}^{(\alpha)}~\textrm{where}~\alpha = 1 \cdots N.
\end{equation*}
Introducing perturbations around the fixed points: $(u_i, v_i) = (u_0, v_0) + (\delta u_i, \delta v_i)$ and substituting into Eq.~\eqref{eq:umatrix} yields linearized equations for $\delta u_i$ and $\delta v_i$,
\begin{equation}
\frac{d \delta u_{i}}{dt} = f_u \delta u_i + f_v \delta v_i + \sum_{j=1}^N L_{ij} \textbf{D} \delta u_j
\end{equation}
\begin{equation}
\frac{dv_{i}}{dt} = g_u \delta u_i + g_v \delta v_i + \sum_{j=1}^N L_{ij} \textbf{D} \delta v_j
\end{equation}
The perturbations can be expanded over the set of Laplacian 
eigenvectors as $\delta u_i (t) = \sum_{\alpha = 1}^{N} c_{\alpha} exp[\lambda_{\alpha} t ] \phi_{i}^{(\alpha)}$ and $\delta v_i (t) = \sum_{\alpha = 1}^{N} c_{\alpha} B_{\alpha} exp[\lambda_{\alpha} t ] \phi_{i}^{(\alpha)}$. Substituting these into equations (A3) and (A4) we obtain the following eigenvalue equation:
\begin{equation}
\lambda_{\alpha}
\begin{pmatrix} 
1   \\ 
B_{\alpha} 
\end{pmatrix} = \begin{pmatrix} 
f_u + \epsilon \Lambda_{\alpha} & f_v \Lambda_{\alpha}   \\ 
g_u  \Lambda_{\alpha} & g_v + \gamma \Lambda_{\alpha}
\end{pmatrix}
\begin{pmatrix}
1 \\
B_{\alpha}
\end{pmatrix}
\end{equation}
The characteristic equation of this system is given by:
\begin{widetext}
\begin{equation}
\lambda_{\alpha}^2 - [Tr(J) + (\epsilon + \gamma) \Lambda_{\alpha}]\lambda_{\alpha} + [\epsilon g_{v} + \gamma  f_{u} ] \Lambda_{\alpha} + Det(J) + \epsilon \gamma \Lambda_{\alpha}^2 = 0,
\end{equation}
\end{widetext}
where $Tr(J) = f_u + g_v$ and 
$Det(J) = f_ug_v - f_vg_{u}$.
\newline
\newline
More compactly this can be written as
\begin{equation}
\lambda_{\alpha}^2 + b(\Lambda_{\alpha})\lambda_{\alpha} + c(\Lambda_{\alpha}) = 0
\label{eq:compact}
\end{equation}
where
\begin{eqnarray}
b(\Lambda_{\alpha}) &=& -[Tr(J) + (\epsilon + \gamma) \Lambda_{\alpha}] \nonumber \\
c(\Lambda_{\alpha}) &=&  \epsilon \sigma\Lambda_{\alpha}^2 + [\epsilon g_{v} + \gamma f_{u}  ] \Lambda_{\alpha} + Det(J).
\end{eqnarray}
The roots of Eq.~\eqref{eq:compact} are
\begin{eqnarray}
\lambda_{\alpha_1}  &=& \frac{-b(\Lambda_{\alpha}) + \sqrt{b(\Lambda_{\alpha})^2 - 4c(\Lambda_{\alpha})}}{2} \nonumber \\
\lambda_{\alpha_2}  &=& \frac{-b(\Lambda_{\alpha}) - \sqrt{b(\Lambda_{\alpha})^2 - 4c(\Lambda_{\alpha})}}{2}.
\end{eqnarray}
\par
In the absence of diffusion, the system is stable and all eigenvalues are negative. Setting all elements of the diffusion matrix to zero yields $b(\Lambda_{\alpha}) = -Tr(J)$ and $c(\Lambda_{\alpha}) = Det(J)$. If both solutions are negative then, $c(\Lambda_{\alpha}) > 0 $ and $-b(\Lambda_{\alpha}) < 0$, thus instability require the following conditions to be satisfied,
\begin{itemize}
\item	$Det(J) = f_ug_v - f_vg_{u} > 0  $

\item $Tr(J) = f_u + g_v < 0 $
\end{itemize}
From the second inequality, we can deduce that $Tr(J) +(\epsilon + \gamma) \Lambda_{\alpha} < f_u + g_v < 0 $ and therefore $\epsilon + \gamma > 0$  (Laplacian eigenvalues are always non-positive). Then, the only positive solution is
\begin{equation}
\lambda_{\alpha}(\Lambda_{\alpha})  = \frac{-b(\Lambda_{\alpha}) + \sqrt{b(\Lambda_{\alpha})^2 - 4c(\Lambda_{\alpha})}}{2}
\label{eq:instability}
\end{equation}
\newline
 In order to induce differentiation, we need $\lambda_{\alpha}(\Lambda_{\alpha}) > 0 $ for some $\alpha$. So, we write: $\frac{-b(\Lambda_{\alpha}) + \sqrt{b(\Lambda_{\alpha})^2 - 4c(\Lambda_{\alpha})}}{2} > 0$ which implies $c(\Lambda_{\alpha}) < 0$. This means that
\newline
\newline
\begin{equation}
c(\Lambda_{\alpha}) = \epsilon \gamma \Lambda_{\alpha}^2 + [\epsilon g_{v} + \gamma f_{u} ] \Lambda_{\alpha} + Det(J) < 0.
\label{eq:calpha}
\end{equation}
Since all Laplacian eigenvalues are  non-positive, we have: $\epsilon g_{v} + f_{u} \gamma > 0 $, and the condition $\epsilon g_{v} + f_{u}\gamma - 4\epsilon \gamma Det(J) > 0$ guarantees negative $c(\Lambda_{\alpha})$ values (we take $\epsilon \gamma > 0$ for an upward opening parabola). The roots of $c(\Lambda_{\alpha})$ are then,

\begin{widetext}
\begin{eqnarray}
\Lambda_{\alpha_1} &=& \frac{-[\epsilon g_{v} + f_{u} \gamma] + \sqrt{[\epsilon g_{v} + \gamma  f_{u} ]^2 - 4\epsilon \gamma Det(J)}}{2 \epsilon \gamma} \nonumber \\
\Lambda_{\alpha_2} &=& \frac{-[\epsilon g_{v} + f_{u} \gamma ] - \sqrt{[\epsilon g_{v} + \gamma  f_{u}  ]^2 - 4 \epsilon \gamma Det(J)}}{2\epsilon \gamma}.
\label{eq:range}
\end{eqnarray}
\end{widetext}

Thus, $c(\Lambda_{\alpha})$ is negative if some of the Laplacian eigenvalues are in the range [$\Lambda_{\alpha_1}$,$\Lambda_{\alpha_2}$] and that guarantees the existence of corresponding growth factors $\lambda_{\alpha}$ to be positive. This is shown in Fig.~\ref{fig:criticality}{\bf b} where the eigenvalues of a PL and ER network are plotted on the curve $c(\Lambda_{\alpha})$. The ER network has all of its eigenvalues (except for $\Lambda_1 = 0$) in the instability range [$\Lambda_{\alpha_1}$,$\Lambda_{\alpha_2}$] whereas the PL network has some of its eigenvalues inside the range and some outside due to its larger degree fluctuation. In Fig.~\ref{fig:criticality}{\bf a,c} we plot the eigenspectrum of both networks.

\end{document}